\documentstyle[12pt,grapcxpc]{article}
\begin{document}

\begin{titlepage}
\normalsize
\begin{center}
{\Large \bf Budker Institute of Nuclear Physics}
\end{center}
\begin{flushright}
BINP 94-88\\
November 1994
\end{flushright}
\vspace{1.0cm}
\begin{center}
{\bf SUPERLUMINAL VELOCITY OF PHOTONS IN A GRAVITATIONAL BACKGROUND}
\end{center}

\vspace{1.0cm}
\begin{center}
{\bf I.B. Khriplovich}\footnote{e-mail address: khriplovich@inp.nsk.su}
\end{center}

\begin{center}
Budker Institute of Nuclear Physics, 630090 Novosibirsk,
Russia
\vspace{4.0cm}
\end{center}

\begin{abstract}
The influence of radiative corrections on the photon propagation in a
gravitational background is investigated without the low-frequency
assumption $\omega \ll m$. The conclusion is made in this way that
the velocity of light can exceed unity.
\end{abstract}

\end{titlepage}

{\bf 1.} The question addressed in the present article was raised
many years ago by Drummond and Hathrell \cite{dh}. They noted that
tidal gravitational forces on the photons, induced by radiative
corrections, in general alter the characteristics of propagation, and
pointed out that due to it photons may travel in some cases at speeds
greater than unity. To be more precise, in a local inertial frame
the induced curvature terms in the Maxwell equation survive and
modify the light cone in different ways for different polarizations.
The analogous conclusion for the neutrino in a Friedman metric was
made somewhat later by Ohkuwa \cite{oh}. Recently the results of
Ref. \cite{dh} were generalized for charged black holes by Daniels and
Shore \cite{ds}.

The approach of Ref. \cite{dh} consisted in expanding the contribution
to the photon effective action from one-loop vacuum polarization to
the lowest order in the inverse electron mass squared $1/m^2$.
Therefore, their result by itself refers strictly speaking to
low-frequency photons with $\omega \ll m$ only. Meanwhile, the velocity
of the wave-front propagation in a dispersive medium is determined by
the asymptotics of the refraction index $n(\omega)$ at $\omega
\rightarrow \infty$ (see, e.g., \cite{le}). It is argued however in
Ref. \cite{dh} that due to the dispersion relation for the refraction
index $n(\omega)$, its high-frequency asymptotics $n(\infty)$ is
related to the low-frequency one $n(0)$ as follows:
\begin{equation}
n(\infty) = n(0) - \frac{2}{\pi}\int_0^{\infty}\frac{d\omega}{\omega}
\;{\rm Im}\;n(\omega).
\end{equation}
Then, since ${\rm Im}\;n(\omega)$ is nonnegative,
$$n(\infty)\leq n(0)$$
which would guarantee the superluminal propagation of the wave front.
The shortcoming of this argument, as pointed out in Ref. \cite{dk}, is
that the sign of ${\rm Im}\;n(\omega)$ in the problem of interest is
not fixed, generally speaking. Indeed, the physical meaning of the
condition
$${\rm Im}\;n(\omega) \geq 0$$
is that in a homogenious medium without instabilities (particle
creation) the wave amplitude can decrease only, due to the loss of
particles from the beam. However, in an inhomogenious medium (and
this is the case of a gravitational background) the processes of
the beam focusing and bunching are possible, leading to the increase
of the wave amplitude which corresponds to
$${\rm Im}\;n(\omega) \leq 0.$$

The analysis performed in Ref. \cite{dh} has demonstrated that even
if the superluminal propagation takes place indeed in this way, it
does not violate causality. Still the effect discussed is quite
unexpected and interesting, and it is certainly worth efforts to find
out whether the predicted phenomenon is a true one or just a result
of an inadequate approximation.  In the present paper the problem is
addressed without the low-frequency assumption $\omega \ll m$. In
this way we come to the conclusion that in a gravitational background
photons can propagate indeed with superluminal velocities.

\bigskip
{\bf 2.} We will start with the discussion of the general structure of
the photon-graviton vertex. To the lowest order in the momenta
$k^{\prime}$ and $k$ of the outgoing and incoming photons respectively,
there is the well-known minimal interaction:
\begin{equation}\label{min}
\frac{1}{2}\kappa h_{\mu\nu}T_{\mu\nu}.
\end{equation}
Here $\kappa h_{\mu\nu}$ is the deviation of the metric from the flat
one:
$$\kappa h_{\mu\nu} = g_{\mu\nu} - \delta_{\mu\nu},\;\;\;\kappa^2
= 32\pi G$$
where $G$ is the Newton constant. The matrix element of
the energy-momentum tensor of electromagnetic field is
\begin{equation}
T_{\mu \nu} = -F^{\prime}_{\mu\lambda}F_{\nu\lambda}
              - F^{\prime}_{\nu\lambda}F_{\mu\lambda}
+ \frac{1}{2}\delta_{\mu\nu}F^{\prime}_{\kappa\lambda}F_{\kappa\lambda}
\end{equation}
where
$$F^{\prime}_{\mu\lambda} = i(k^{\prime}_{\mu} e^{\prime}_{\lambda}
                                -k^{\prime}_{\lambda} e^{\prime}_{\mu}),$$
$$F_{\nu\lambda} = -i(k_{\nu} e_{\lambda} - k_{\lambda} e_{\nu}),$$
and $e^{\prime}_{\mu},\;e_{\nu}$ are the polarization vectors of the
outgoing and incoming photons, respectively.

To investigate other possible structures for this vertex, let us
introduce vectors
$$p_{\mu} = k^{\prime}_{\mu} + k_{\mu},\;\;q_{\mu} = k^{\prime}_{\mu}-k_{\mu}$$
orthogonal on mass shell, $p_{\mu}q_{\mu} =0$. The next two
independent vectors can be conveniently chosen as
$$g_{1\mu} = k^{\prime}_{\nu} F_{\nu\mu},\;\;
  g_{2\mu} = k_{\nu} F^{\prime}_{\nu\mu}.$$
Then the general vertex contains beside (\ref{min}) three other
symmetric second-rank tensors, bilinear in $F^{\prime},\;F$ and
orthogonal to $q$:
\begin{equation}
\tau_{1\mu\nu} = p_{\mu} p_{\nu}F^{\prime}_{\kappa\lambda}F_{\kappa\lambda},
\end{equation}
\begin{equation}
\tau_{2\mu\nu} = (q^2\delta_{\mu\nu} - q_{\mu}q_{\nu})
                  F^{\prime}_{\kappa\lambda}F_{\kappa\lambda},
\end{equation}
\begin{equation}
\tau_{3\mu\nu} = g_{1\mu}g_{2\nu} + g_{1\nu}g_{2\mu}
= k^{\prime}_{\alpha} F_{\alpha\mu} k_{\beta} F^{\prime}_{\beta\nu}
  + k^{\prime}_{\alpha} F_{\alpha\nu} k_{\beta} F^{\prime}_{\beta\mu};
\end{equation}
$$q_{\mu}\tau_{i\mu\nu} = 0.$$

The interaction of $\tau_{2,3}$ with an external gravitational field
can be immediately rewritten in a covariant form:
\begin{equation}
\kappa h_{\mu\nu}\tau_{2\mu\nu}
                           = R\;F^{\prime}_{\kappa\lambda}F_{\kappa\lambda},
\end{equation}
\begin{equation}
\kappa h_{\mu\nu}\tau_{3\mu\nu}
 = - R_{\mu\nu\kappa\lambda} F^{\prime}_{\mu\nu}F_{\kappa\lambda}
\end{equation}
where $R$ and $R_{\mu\nu\kappa\lambda}$ are the scalar curvature and
the Riemann tensor, respectively.

As to $\tau_{1\mu\nu}$, its interaction reduces to
\begin{equation}\label{an}
\kappa h_{\mu\nu}\tau_{1\mu\nu}
 = 4R_{\mu\nu\kappa\lambda} F^{\prime}_{\mu\nu}F_{\kappa\lambda}
  - R F^{\prime}_{\kappa\lambda}F_{\kappa\lambda}
  + 2q^2 \kappa h_{\mu\nu} T_{\mu\nu}.
\end{equation}
Let us note also that as reducible in this sense is the widely used covariant
structure with the Ricci tensor $R_{\mu\nu}$:
\begin{equation}\label{ri}
R_{\mu\nu} F^{\prime}_{\mu\lambda}F_{\nu\lambda}
  = \frac{1}{4} R F^{\prime}_{\kappa\lambda}F_{\kappa\lambda}
    - \frac{1}{4} q^2 \kappa h_{\mu\nu} T_{\mu\nu}.
\end{equation}
Thus, the most general tensor structure for the photon-graviton
vertex, valid at any frequencies and momentum transfers, can be
presented as
\begin{equation}\label{gen}
\frac{1}{2}\kappa h_{\mu\nu}T_{\mu\nu}f_1(q^2)
+ R_{\mu\nu\kappa\lambda} F^{\prime}_{\mu\nu}F_{\kappa\lambda}f_2(q^2)
+ R F^{\prime}_{\kappa\lambda}F_{\kappa\lambda}f_3(q^2).
\end{equation}
The lowest order QED contribution to the form-factors $f_i$ was
calculated in Refs. \cite{bg,mi}. The first nontrivial terms of their
expansion in $q^2$ are (see \cite{dh}):
\begin{equation}
f_1 = 1 + \frac{11\alpha}{720\pi} \frac{q^2}{m^2},\;\;
f_2 = - \frac{\alpha}{360\pi m^2},\;\;
f_3 = - \frac{\alpha}{144\pi m^2}.
\end{equation}

Similar analysis can be performed for the neutrino-graviton
interaction. As to the structure with $R_{\mu\nu\kappa\lambda}$, it
is kinematically impossible at all for a spin 1/2 particle. The
interaction with the scalar curvature is forbidden for a
two-component neutrino by helicity arguments. Therefore, the
neutrino-graviton vertex is reduced effectively to interaction
(\ref{min}) with a form-factor.  An identity analogous to (\ref{ri})
allows one to use an alternative form: the pointlike minimal
interaction of the energy-momentum tensor with $h_{\mu\nu}$ plus the
interaction of the same energy-momentum tensor with $R_{\mu\nu}$.
Just this last form was used in Ref. \cite{oh}.

\bigskip
{\bf 3.} Passing over at last to the photon propagation problem, let
us emphasize that the form-factors in amplitude (\ref{gen}) depend on
the momentum transfer only, but not on the photon energy itself.  Of
course, this property is in no way confined to the lowest order loop
calculated in Refs. \cite{bg,mi}, but refers to a general vertex with
two on-mass-shell particles.  Moreover, when light propagates in a
gravitational field of a macroscopic length scale $L$, the typical
impact parameters $\sim L$ are large as compared to the Compton
wave-length $m^{-1}$ (or any other dimensional parameter possibly
involved in the radiative corrections), and therefore one can confine
to the values of the form-factors $f_i$ at $q^2 = 0$.

The lowest order correction discussed modifies the Maxwell equation
in the region where $R_{\mu\nu} = R = 0$, and at $\omega L \gg 1$, as
follows \cite{dh}:
\begin{equation}\label{max}
D_{\mu}F^{\mu\nu} + \xi R^{\mu\nu}_{\rho\tau} D_{\mu}F^{\rho\tau} = 0;\;\;
\xi = \frac{\alpha}{90\pi m^2}.
\end{equation}
The structure $\xi R^{\mu\nu}_{\rho\tau}$ in this expression can be
considered obviously as an anisotropic contribution to a refraction
index which in general leads to a superluminal photon velocity.

However, the photon interaction with a gravitational background,
induced by radiative corrections, certainly contains terms of higher
order in curvature. How will they influence the photon propagation?

As distinct from the three-particle vertex discussed, the diagrams
generating the terms nonlinear in $R$ (from now on $R$ is a
generic notation for $R_{\mu\nu\kappa\lambda},\;R_{\mu\nu},\; R$)
have more external lines and therefore certainly depend on the photon
energy. But by dimensional reasons it is quite natural to expect that
it is $\;R/\omega^2\;$ which serves as a parameter for the
high-frequency

\pagebreak

\noindent expansion of the photon-background interaction
\footnote{By the way, this is exactly what happens when one
considers the effect of the curvature tidal forces on the propagation
of a finite wave packet in classical gravity \cite{dk}. Even to
linear approximation the effective refraction index for a photon of
the helicity $\lambda$ propagating along the $z$ axis in a space with
vanishing $R_{\mu\nu}$, reduces in a local inertial frame to
$$n = 1 + \frac{1}{2\omega^2}(R_{1212} + i\lambda R_{1230}).$$}. Such
a behaviour in the high-frequency limit is much more natural than the
expansion in $R/m^2$ with mass singularities in the asymptotic
region. In this sense the photon-graviton vertex is an exception:
being $\omega$-independent (kind of a subtraction constant in the
dispersion relation), it has no choice at $q^2 \ll m^2$ but generate
linear terms of the type $R/m^2$. Thus the terms nonlinear in $R$ die
out at $\omega \rightarrow \infty$ and do not influence the
wave-front propagation. On the other hand, there is a case when those
nonlinear terms are certainly inessential at any frequency: that of a
weak gravitational background.

Since now the low-frequency limit can be abandoned, the situation
changes as well with the problem whether the phenomenon of
superluminal propagation is observable, at least in principle. The
difficulty pointed out in Refs. \cite{dh,ds} is as follows. If the
curvature length scale is $L\;(R \sim L^{-2})$, then according to
Eq.(\ref{max}) the velocity shift caused by the radiative
correction is
\begin{equation}
\delta v \sim \frac{\alpha}{m^2 L^2}.
\end{equation}
The time available for examining signals is also $L$. So, the
corresponding position discrepancy of a signal constitutes
\begin{equation}
\delta s \sim L\delta v \sim \frac{\alpha}{m^2 L}\ll \frac{1}{m}.
\end{equation}
It is not exactly clear how such a distance can be resolved with
frequencies $\omega \ll m$ discussed in Refs. \cite{dh,ds}. Of course,
going beyond the low-frequency approximation removes this difficulty
in principle.

The arguments presented in this work give strong reasons to believe
that the effect of superluminal propagation of photons in a
gravitational background does exist.

\bigskip
I am grateful to A.D. Dolgov, V.M. Khatsymovsky and V.V. Sokolov for
discussions.
\vspace{5.5cm}

\pagebreak

\end{document}